# Steepness and spectrum of nonlinear deformed shallow water wave

Narcisse Zahibo[1], Irina Didenkulova[2,3], Andrey Kurkin[2] and Efim Pelinovsky[2,3]

[1] Physics Department, University of Antilles Guyane, Pointe-a-Pitre, Guadeloupe, France

[2] Applied Mathematics Department, State Technical University, Nizhny Novgorod, Russia

[3] Department of Nonlinear Geophysical Processes, Institute of Applied Physics, Nizhny Novgorod, Russia

Process of the nonlinear deformation of the shallow water wave in a basin of constant depth is studied. The characteristics of the first breaking are analyzed in details. The Fourier spectrum and steepness of the nonlinear wave is calculated. It is shown that spectral amplitudes can be expressed through the wave front steepness, and this can be used for practical estimations.

Corresponding author: Professor Efim Pelinovsky
Department of Nonlinear Geophysical Processes,
Institute of Applied Physics,
Russian Academy of Sciences,
46 Uljanov Street, Nizhny Novgorod, Russia
Tel: 007-8312-164839
Fax: 007-8312-365976
Email: pelinovsky@hydro.appl.sci-nnov.ru

# 1. Introduction

The process of the nonlinear wave evolution in the shallow water resulting to the wave breaking is well-known, and it can be described analytically in the framework of the nonlinear shallow-water theory (see, for instance, books: Stoker, 1957: Whitham, 1974; Engelbrecht et al, 1988; Voltsinger et al, 1989; Arseniev et al, 1991; Tan, 1992). Mathematically, the wave breaking can be considered as the crossing of the characteristics of the hyperbolic system of the shallow water (gradient catastrophe). A lot of observations of the wave breaking and its transformation into the undular bore were made during the huge tsunami in the Indian Ocean occurred on 26[th] December 2004 after the earthquake with magnitude 9.3. Figure 1 shows the set of photos of tsunami approaching to the coast made by the Canadian couple (these photos have been very often demonstrated in TV after the event). The increasing of the steepness of the tsunami wave front is obtained also in numerical simulation of the tsunami wave propagation on long distances (Zahibo et al, 2006) and predicted theoretically (Hammack, 1973; Ostrovsky and Pelinovsky, 1976; Murty, 1977; Pelinovsky, 1982). The same processes are observed when sea waves entry in river mouths (Pelinovsky, 1982; Tsuji et al, 1991) and straits or channels (Pelinovsky and Troshina, 1994; Wu and Tian, 2000; Caputo and Stepanyants, 2003). Meanwhile, we do not know publications where the characteristics of the nonlinear deformed wave such as steepness, spectrum and location of breaking point have been analyzed in details. Here we would like to consider the nonlinear deformation of the shallow water wave in a basin of constant depth with no limitation on the wave amplitude. The spatial evolution of the nonlinear deformed wave and the characteristics of the first breaking are analyzed in section 2. The wave steepness and the Fourier spectrum of the nonlinear deformed periodic wave are studied in section 3. Obtained results are summarized in Conclusion.

## 2. Spatial evolution of the shallow water wave

The basic equations of the nonlinear shallow water theory can be written in the form (Stoker, 1955)

$$\frac{\partial u}{\partial t} + u\frac{\partial u}{\partial x} + g\frac{\partial \eta}{\partial x} = 0,$$

$$\frac{\partial \eta}{\partial t} + \frac{\partial}{\partial x}[(h+\eta)u] = 0,$$

(1)

where $\eta$ is the water level displacement, $u$ is the horizontal velocity of water flow, $g$ is a gravity acceleration and $h$ is unperturbed water depth assumed to be constant.

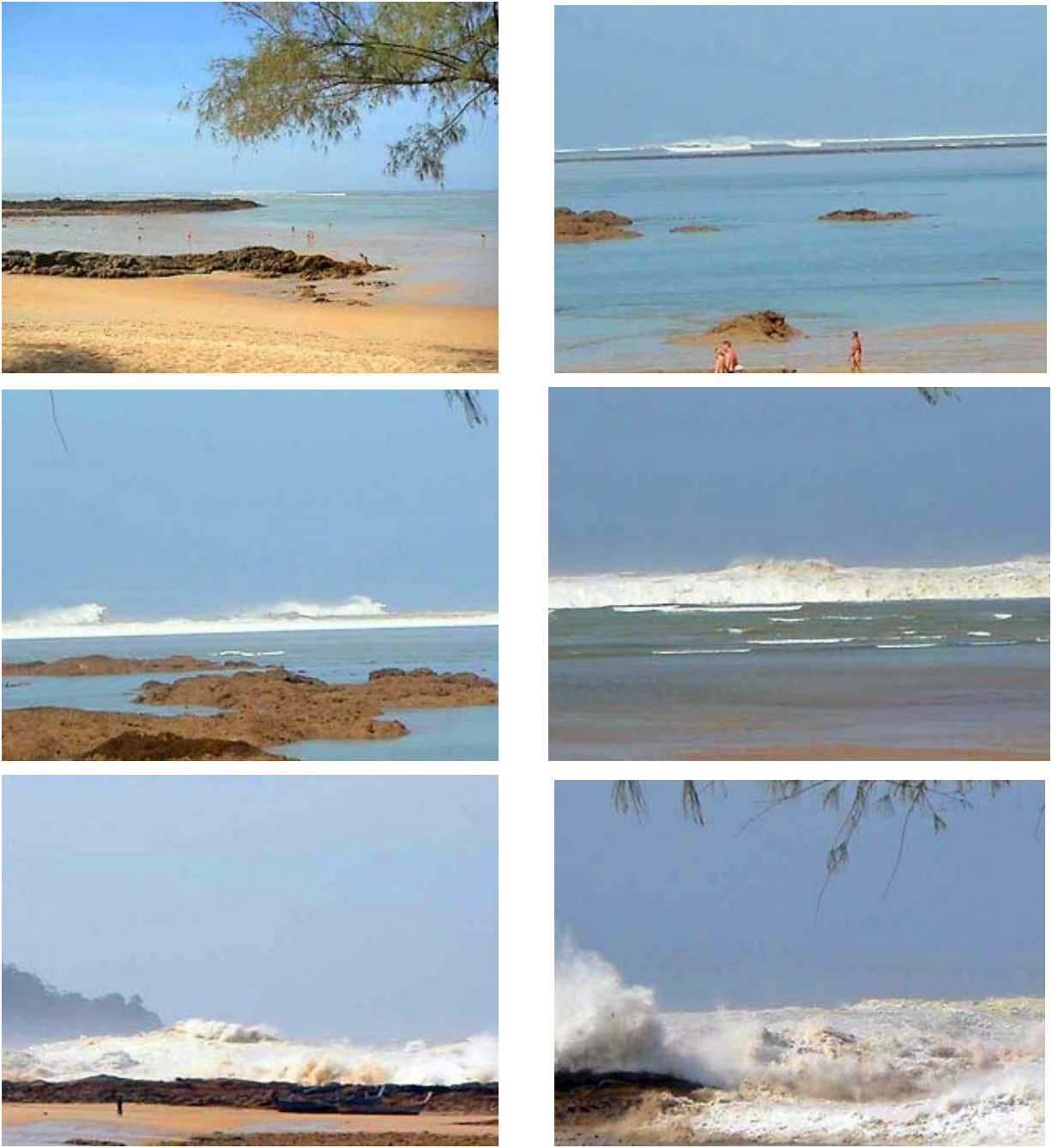

**Fig. 1.** Snapshot of tsunami approaching to the coast (Indian Ocean, December 26, 2004)

In the unidirectional wave the flow velocity depends on the water displacement only, and after substitution $u = u(\eta)$ in (1), this relation can be found explicitly

$$u = 2\left(\sqrt{g(h+\eta)} - \sqrt{gh}\right) \quad (2)$$

(for definition, we consider waves propagated in direction, $x > 0$). At this condition, the system (1) is reduced to the first-order quasi-linear partial differential equation (Whitham, 1974; Voltsinger et al, 1989)

$$\frac{\partial \eta}{\partial t} + V(\eta)\frac{\partial \eta}{\partial x} = 0, \qquad (3)$$

where the wave speed (characteristic speed), *V* is

$$V(\eta) = 3\sqrt{g(h+\eta)} - 2\sqrt{gh}. \qquad (4)$$

It is important to mention that the equation (3) is an exact equation and valid formally for the waves of any amplitude if the effects of dispersion and dissipation are neglected.

We will solve the equation (3) with the boundary condition $\eta(t,x=0) = \eta_0(t)$, which is typical for the cases when the wave is generated in the laboratory tank by the wavemaker. Thus, such solution (it is called as simple or Riemann wave in nonlinear acoustics, see, Engelbrecht et al, 1988; Gurbatov et al, 1991) is

$$\eta(x,t) = \eta_0\left(t - \frac{x}{V(\eta)}\right), \qquad \text{or} \qquad t - \frac{x}{V(\eta)} = \tau(\eta), \qquad (5)$$

where $\tau(\eta)$ is an inverse function to $\eta_0(t)$ and it is determined by the wavemaker. First of all, it is important to mention that the wave may propagate from the wavemaker in direction $x > 0$ if its speed, *V* is positive corresponding that

$$\eta > -\frac{5}{9}h, \qquad (6)$$

and, therefore, wave through should be not deep.

The implicit formula (5) describes the simple or Riemann wave well known in the nonlinear acoustics (Rudenko and Soluyan, 1977; Engelbrecht et al, 1988; Gurbatov et al, 1991). This solution describes the nonlinear deformation of the wave with distance; the steepness of its face slope increases with distance. The temporal derivative of the wave profile can be calculated in the explicit form

$$\frac{\partial \eta}{\partial t} = \frac{d\eta_0/d\tau}{1 + x dV^{-1}(\eta_0)/d\tau}. \qquad (7)$$

On the face slope ($\partial \eta/\partial t > 0$) the temporal derivative $dV^{-1}/dt$ is negative, and the denominator in (6) decreases with distance; the temporal derivative of the wave profile increases and tends to infinity at distance $x = X$. The breaking (nonlinear) length characterized the first breaking equals to

$$X = \frac{1}{\max(-dV^{-1}/dt)}. \tag{8}$$

Therefore, the wave begins to break at the point on the wave profile where the inverse speed has the maximum derivative, and this point in general does not coincide with the point of the wave profile with the maximum steepness. As a detailed example, the initial sine shape of the generated wave will be analyzed. Such a wave, $\eta_0(t) = a \sin(\omega t)$, has the maximum derivative $a\omega$ in the wave point with the zero displacement of the water level. The breaking begins in the trough, and the phase ($\theta = \omega t_*$) and displacement ($\zeta = \eta_*/h$, $\eta_* = a \sin\theta$) of the breaking point (see Figure 2 for definitions) depend from the wave amplitude ($A = a/h$) through the algebraic dimensionless expressions

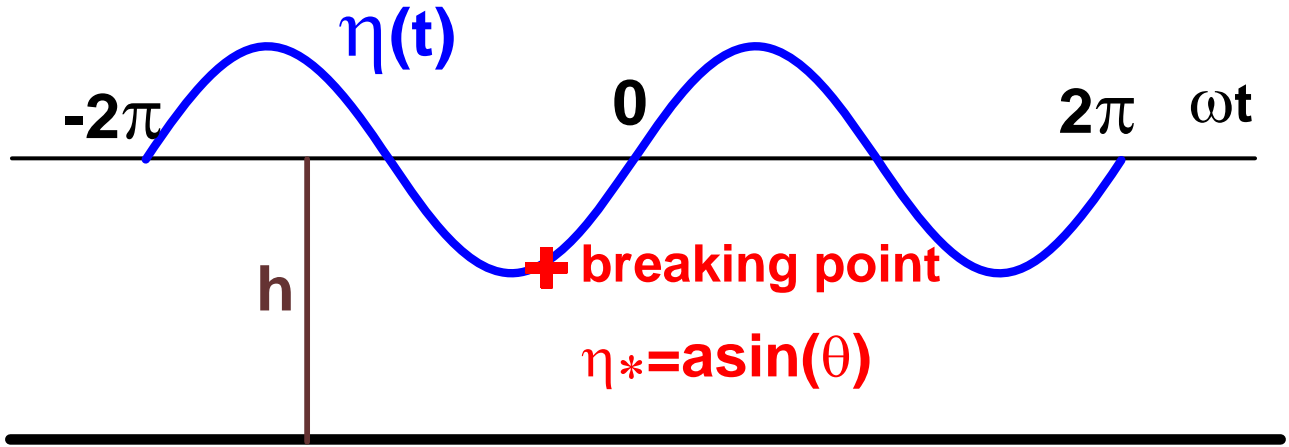

**Fig. 2.** Location of the breaking point in trough on the shallow water wave

$$A = \sqrt{\frac{\zeta^2(3\sqrt{1+\zeta}+2)-2\zeta(3\sqrt{1+\zeta}-2)}{9\sqrt{1+\zeta}-2}},$$

$$\theta = \arcsin\left(\frac{\zeta}{A}\right). \tag{9}$$

Both characteristics are shown in Figures 3 and 4. At small wave amplitudes the first breaking is appeared in the wave on the zero level (unperturbed fluid surface), and here the following asymptotic formulas can be used to estimate displacement and phase of the breaking point

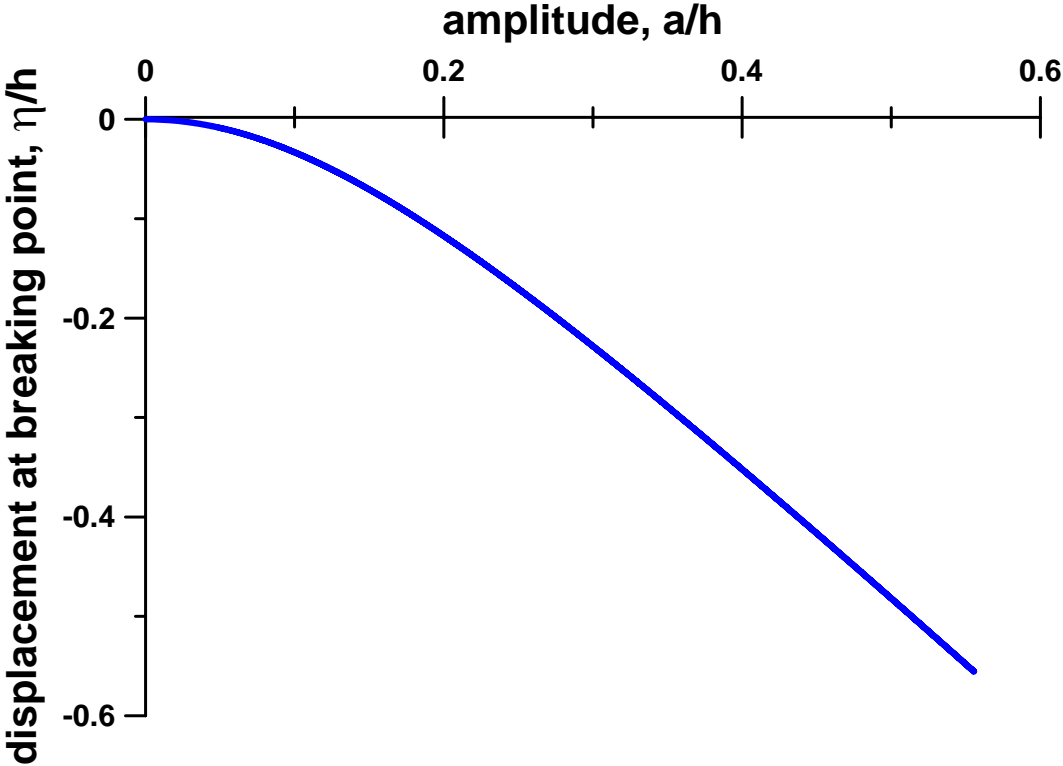

**Fig. 3.** Displacement at the breaking point versus the wave amplitude

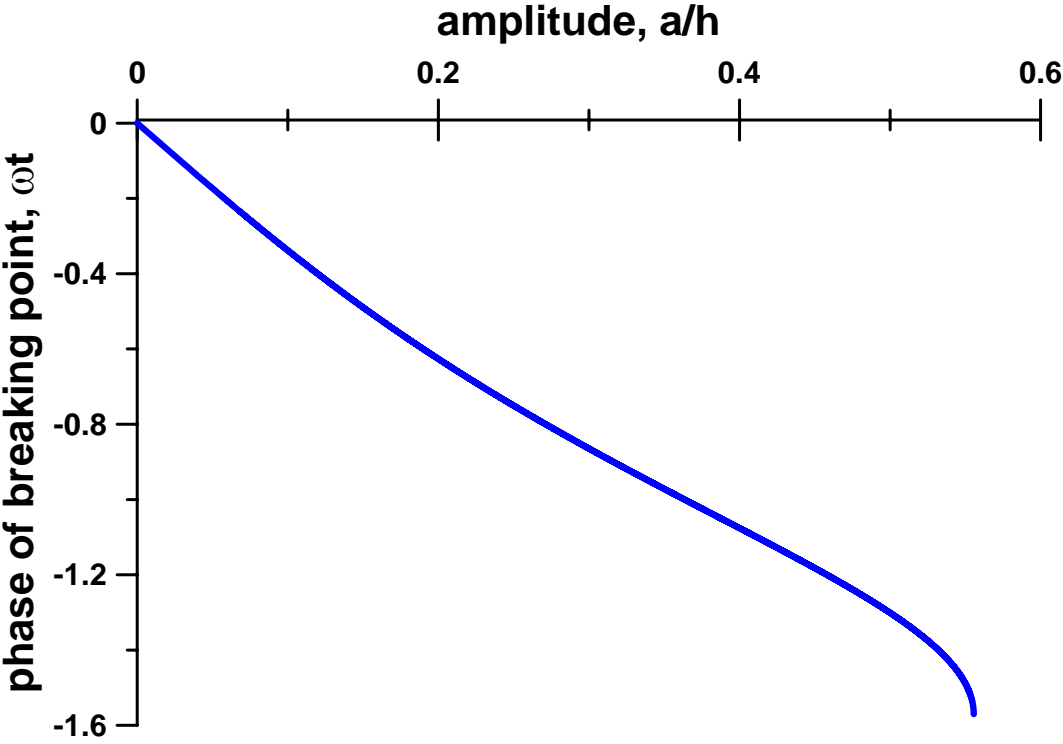

**Fig. 4.** Phase of the breaking point versus the wave amplitude

$$\zeta \approx -\frac{7}{2}A^2, \qquad \theta \approx -\frac{7}{2}A. \qquad (10)$$

When the wave amplitude tends to the critical value (6), the breaking point is shifting in the end of the wave trough

$$\zeta \approx -A = \frac{5}{9}, \qquad \theta \approx -\frac{\pi}{2}. \qquad (11)$$

The distance when the first breaking is appeared (breaking length) can be found from (8)

$$\omega T = \frac{2\sqrt{1+\zeta}\,(3\sqrt{1+\zeta}-2)^2}{3A\cos\theta}, \qquad (12)$$

where the expressions (9) should be used. In fact, we introduce the breaking time, $T=X/(gh)^{1/2}$ reducing the number of the dimension quantities. The breaking time decreases when the wave amplitude increases (Figure 5), tends to zero at wave amplitude tends to the critical value 5/9. Thus, the wave of large amplitude breaks near the wavemaker and in fact does not propagate. For weak amplitude waves the breaking time is high, and it can be described by the asymptotic formula

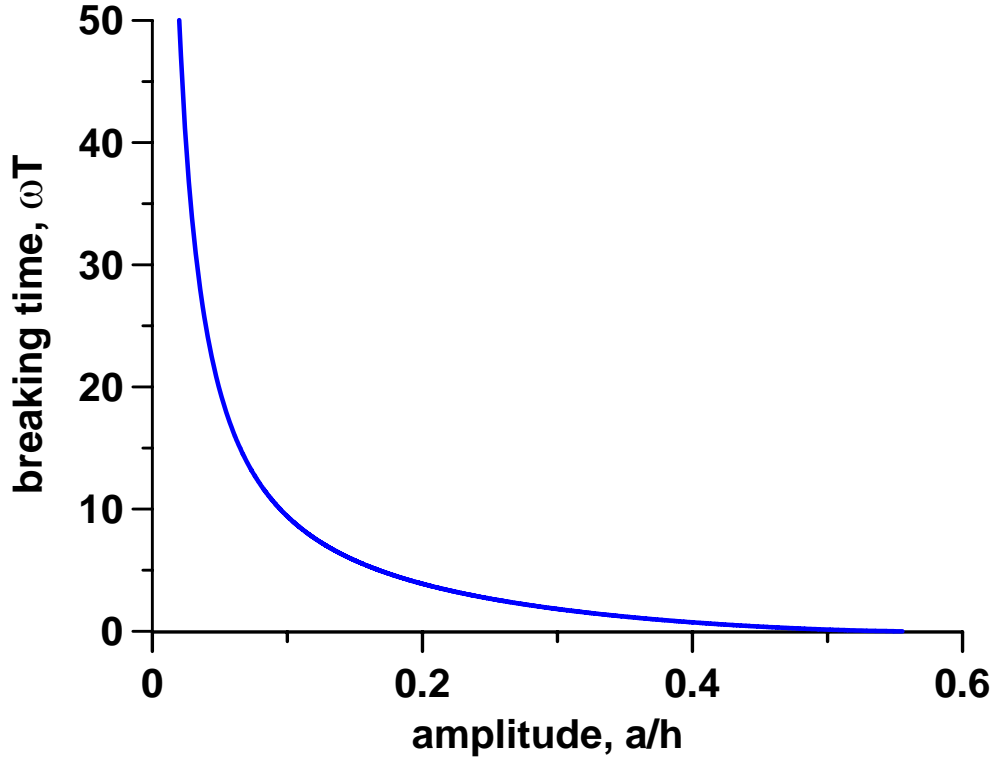

**Fig. 5.** Breaking time versus the wave amplitude

$$\omega T \approx \frac{2}{3A}. \qquad (13)$$

Taking into the account its importance for practice let us give this formula in dimension form

$$X = \frac{2\sqrt{gh}\,h}{3\omega a} = \frac{\lambda h}{3\pi a}, \qquad (14)$$

where the wavelength, λ is determined by the linear dispersion relation of long waves

$$\lambda = \frac{2\pi\sqrt{gh}}{\omega}. \qquad (15)$$

The weak amplitude wave has to run the long distance of many wavelengths while the nonlinear effects will be significant and the wave breaks.

## 3. The wave steepness and Fourier spectrum

The wave steepness is the measured characteristics of the wave field important for applications, and it can be calculated exactly from (5)

$$\frac{\partial \eta}{\partial x} = -\frac{1}{V}\frac{\partial \eta}{\partial t} = -\frac{V^{-1}(\eta_0)d\eta_0/d\tau}{1 + x dV^{-1}(\eta_0)/d\tau}. \qquad (16)$$

The maximum steepness is achieved at the point of wave profile where $V^1$ has the maximum temporal derivative; the first breaking occurs at this point. For the maximum steepness from (16) taking into the account the definition of the breaking length (8) follows

$$s = \max(\partial \eta / \partial x) = \frac{s_0}{1 - x/X}, \qquad (17)$$

where $s_0 = \partial\eta_0/\partial x = V^{1}\partial\eta_0/\partial\tau$ is the initial steepness of the wave in the point $\eta_*$. The maximum steepness increases very rapidly in the vicinity of the breaking point, see Figure 6. The minimum steepness is achieved on the back face of the wave, and it is varied with distance as

$$s_{min} = \min(\partial \eta / \partial x) = \frac{s_0}{1 + x/X}. \tag{18}$$

The minimum steepness reduces with distance and it is twice less then the initial steepness at the breaking point. Similarly the steepness of the nonlinear deformed wave can be calculated for other points on the wave profile.

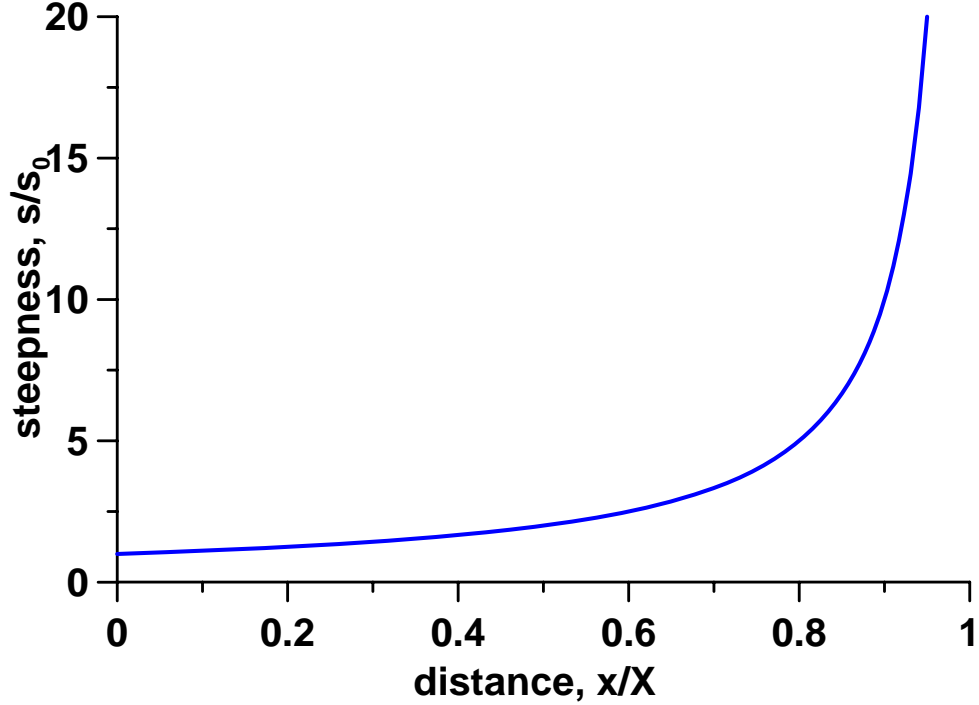

**Fig. 6.** The wave steepness versus the distance from wave maker

For practice it is important to know the frequency spectrum of the wave field. In general form the Fourier integral can be written in the explicit form (Pelinovsky, 1976)

$$S(\omega) = \int \eta(x,t) \exp(-i\omega t) dt = \frac{1}{i\omega} \int \frac{d\eta_0}{d\tau} \exp(-i\omega[\tau + x/V(\eta_0)]) d\tau. \tag{19}$$

It is impossible to calculate this integral even for the monochromatic initial disturbances. Let us consider here the case of weak, but finite amplitudes when the wave propagates on long distances with no breaking. Using the Taylor's series for inverse velocity $V(\eta)$ as $V^{-1}(\eta) = (gh)^{-1/2}(1-3\eta/2h)$, the integral (19) can be calculated exactly and the wave field of any distance from the wave maker is

$$\eta(t,x) = \sum_{n=1}^{\infty} A_n(x) \sin(n\omega[t - x/\sqrt{gh}]) = \frac{4h\sqrt{gh}}{3\omega x} \sum_{n=1}^{\infty} \frac{1}{n} J_n\left(\frac{3n\omega x a}{2h\sqrt{gh}}\right) \sin(n\omega[t - x/\sqrt{gh}]), \tag{20}$$

where $J_n$ is the Bessel function of $n$-order. Spectral amplitudes can be re-written in the following form using the same accuracy for the breaking length (14)

$$A_n(x) = 2a \frac{X}{nx} J_n\left(\frac{nx}{X}\right). \qquad (21)$$

The amplitudes of high harmonics grow with distance from the wave maker, and the amplitude of basic harmonics decreases because the energy transfers to the high harmonics (Figure 7). It is important to mention that harmonic amplitudes even at the breaking point are relative weak and decrease with increase of harmonic number.

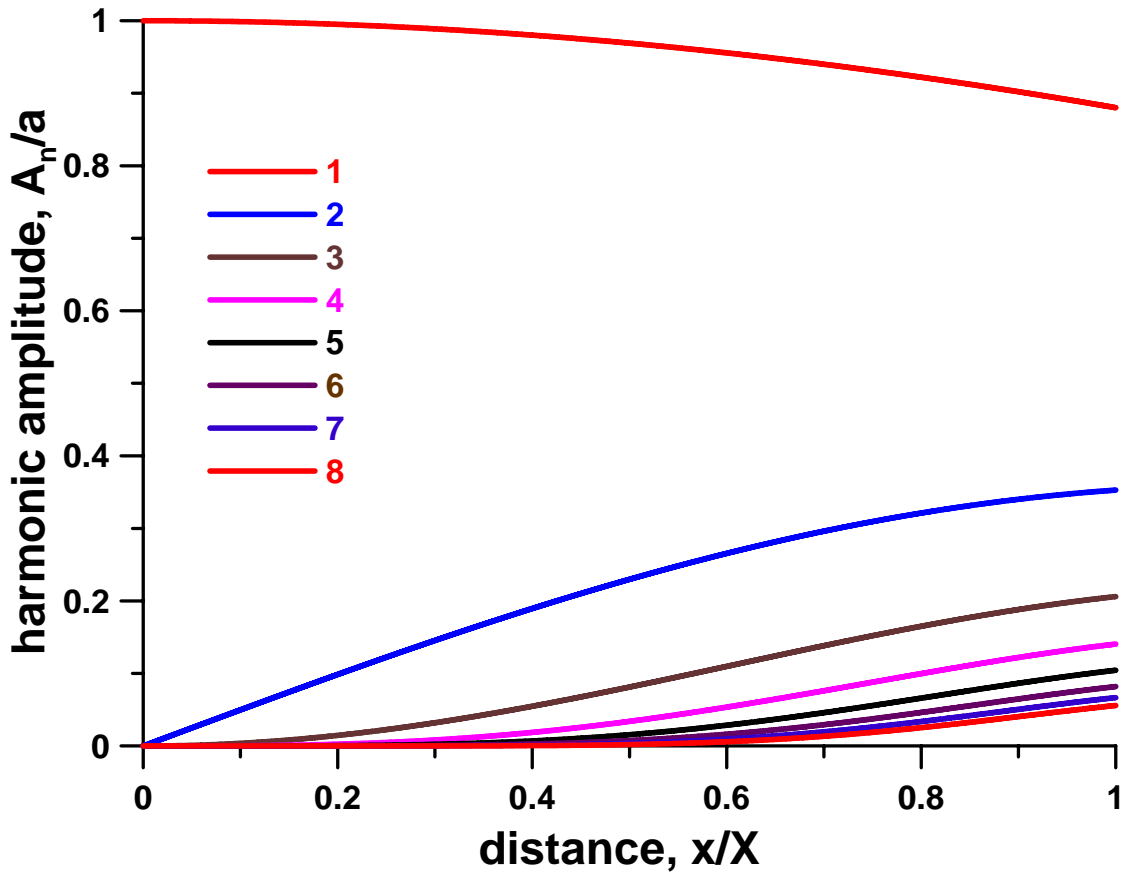

**Fig. 7.** Harmonic amplitudes versus the distance from the wave maker (numbers harmonics are displayed)

In oceanic conditions the distance from the wave source is unknown. More productive is to have the relation between the harmonic amplitudes and the wave steepness. Using (17) for maximum steepness, the formula (21) can be written as

$$A_n(s) = \frac{2a}{n(1-s_0/s)} J_n\left(n\left[1-\frac{s_0}{s}\right]\right). \tag{22}$$

The relation between the harmonic amplitudes and wave steepness is displayed in Figure 8. As it can be seen, for large values of the wave steepness the harmonic amplitudes tend to the constant values. This limited spectrum of the shallow water wave does not depend from the initial wave steepness

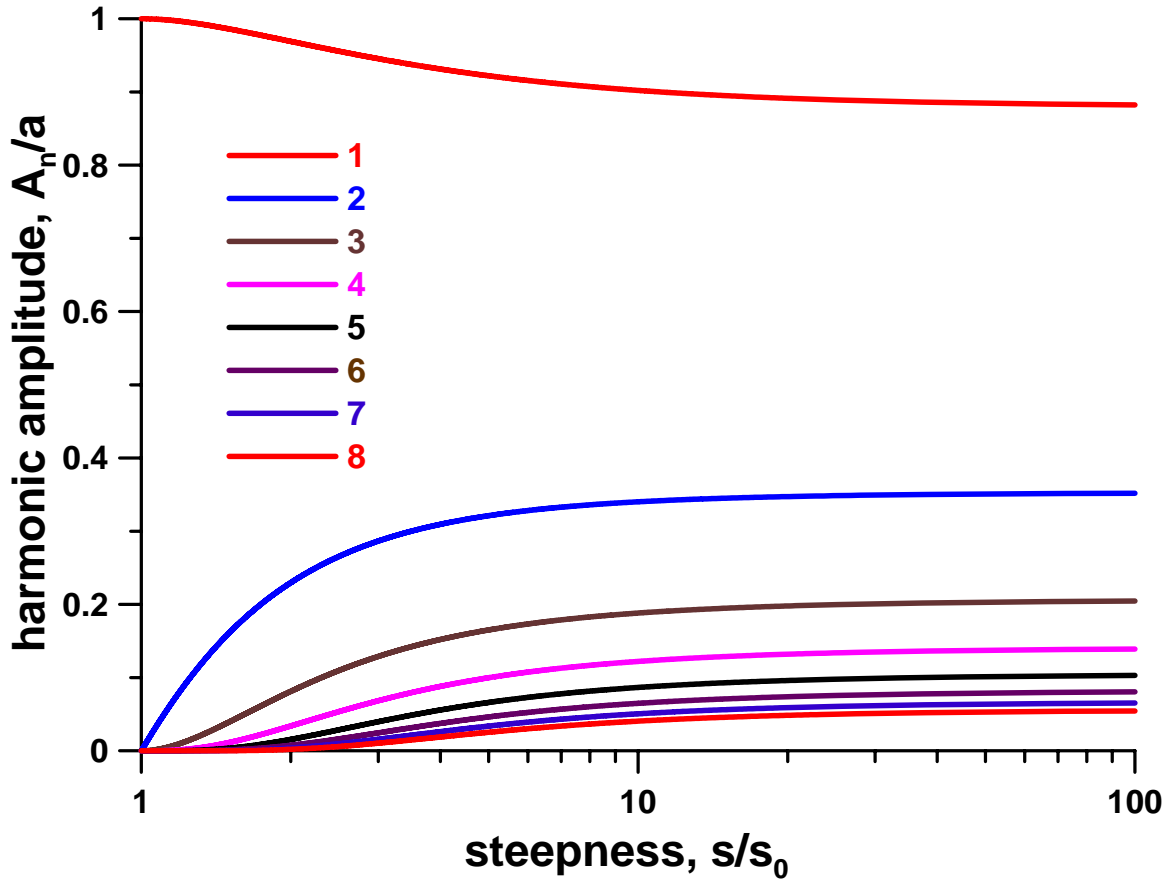

**Fig. 8.** Harmonic amplitudes versus maximum wave steepness

$$\overline{A}_n = \frac{2a}{n} J_n(n), \tag{23}$$

it is shown in Figure 9. This function is very well approximated by the power asymptotic, $n^{-1.3}$, presented in Figure 9 by solid line. Theoretically, the spectrum of the nonlinear nondispersive wave field has been studied in details in (Gurbatov et al, 1991). According to the theory, the asymptotic $\omega^{-3/2}$ appears in the vicinity of the breaking point and this is close to the approximated curve (23) calculated for the periodic waves. After the wave breaking, the asymptotic $\omega^{-1}$ in high-frequency range is forming; this corresponds to the "jump" functions.

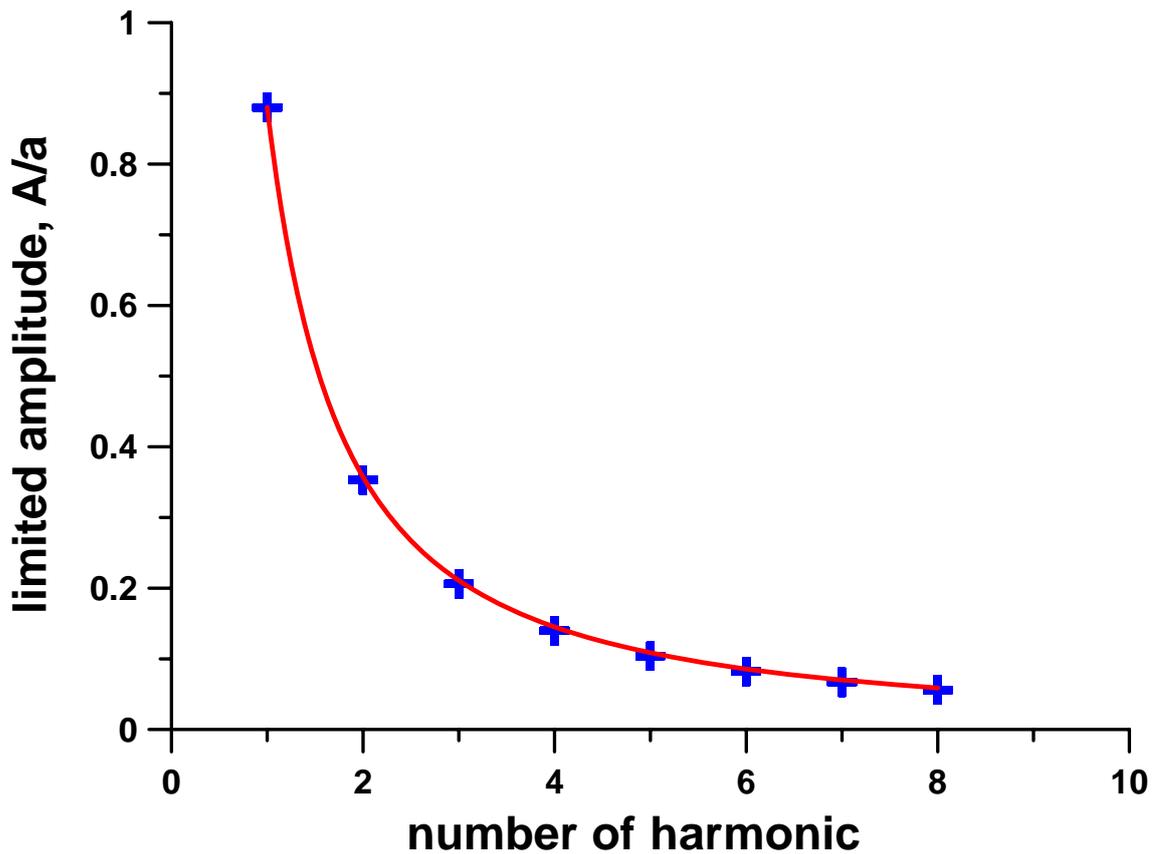

Fig. 9. Limited spectrum of the shallow water wave in the vicinity of the breaking point (solid line – asymptotic $n^{-1.3}$)

## 4. Conclusion

The behavior of the nonlinear shallow-water wave generated by the wavemaker is discussed in the framework of the exact solution in the form of the Riemann (simple) wave. It is shown that the initial sine wave can propagate as the smooth wave only if its amplitude satisfies to the condition, *a < 5h/9*, where *h* is water depth. The wave begins to break at the point on the wave profile where the local value of the inverse velocity of propagation is maximal. The breaking length is calculated; it decreases when the wave amplitude increases. The steepness and the spectrum of the nonlinear deformed wave are calculated in the explicit form. The spectral amplitudes of wave harmonics can be expressed through the local value of the maximum steepness of wave front. The Fourier spectrum has the universal shape for the very steep waves. These estimates of the wave spectrum can be used in the engineering practice.

This study is supported in particular by the grants from INTAS (03-51-4286 and 05-96-4309) and RFBR (05-05-64265 and 06-05-91553).